\documentclass[paper=A4,11pt,DIV=12,headings=big]{scrartcl}
\pdfoutput=1

\usepackage[a4paper, top=1.2in, bottom=1.2in, left=1.1in, right=1.1in]{geometry}
\usepackage[utf8]{inputenc}
\usepackage{amsfonts}
\usepackage{amsmath,amssymb,slashed}
\usepackage{mathrsfs}
\usepackage{mathtools}

\usepackage[normal]{caption}

\usepackage{titlesec}

\titleformat{\section}
{\Large \sffamily \bfseries \boldmath}{\thesection}{0.5em}{}
\titleformat{\subsection}
{\large \sffamily \bfseries \boldmath}{\thesubsection}{0.5em}{}
\titleformat{\subsubsection}
{\sffamily \bfseries \boldmath}{\thesubsubsection}{0.5em}{}

\usepackage{cancel}
\usepackage[dvipsnames]{xcolor}
\usepackage{cite,bbm}
\usepackage{enumerate}
\usepackage{graphicx}
\graphicspath{{./Figs/}}
\usepackage{wrapfig}

\usepackage{bbold}
\usepackage{multirow}

\renewenvironment{abstract}{%
\begin{minipage}{0.95\textwidth}
}
{\par\noindent\end{minipage}}

\usepackage[multiple]{footmisc}
\let\oldfootnote\footnote\renewcommand\footnote[1]{\oldfootnote{\hspace{2mm}#1}}

\definecolor{darkblue}{rgb}{0,0,0.9}
\usepackage{hyperref}
\hypersetup{
linktocpage,
colorlinks,
citecolor=darkblue,
filecolor=darkblue,
linkcolor=darkblue,
urlcolor=darkblue
}

\allowdisplaybreaks
\addtokomafont{disposition}{\rmfamily\boldmath}

\newcommand{\mc}{\mathcal}

\newcommand{\<}{\langle}
\renewcommand{\>}{\rangle}

\newcommand{\mcX}{\mc X}
\newcommand{\EHC}{{\rm EHC}}
\newcommand{\HC}{{\rm HC}}

\newcommand{\lab}{{\rm lab}}
\newcommand{\ann}{{\rm ann}}
\newcommand{\NGB}{{\rm NGB}}
\newcommand{\HB}{{\rm HB}}

\def\sla#1{\setbox0=\hbox{$#1$}\dimen0=\wd0
      \setbox1=\hbox{/} \dimen1=\wd1 \ifdim\dimen0>\dimen1
      \rlap{\hbox to \dimen0{\hfil/\hfil}} #1                        \else
      \rlap{\hbox to \dimen1{\hfil$#1$\hfil}}
      /   \fi}

\newcommand{\be}{\begin{equation}}
\newcommand{\ee}{\end{equation}}
\newcommand{\bea}{\begin{eqnarray}}
\newcommand{\eea}{\end{eqnarray}}

\newcommand{\nn}{\nonumber}

\DeclareOldFontCommand{\rm}{\normalfont\rmfamily}{\mathrm}
\DeclareOldFontCommand{\sf}{\normalfont\sffamily}{\mathsf}
\DeclareOldFontCommand{\tt}{\normalfont\ttfamily}{\mathtt}
\DeclareOldFontCommand{\bf}{\normalfont\bfseries}{\mathbf}
\DeclareOldFontCommand{\it}{\normalfont\itshape}{\mathit}
\DeclareOldFontCommand{\sl}{\normalfont\slshape}{\@nomath\sl}
\DeclareOldFontCommand{\sc}{\normalfont\scshape}{\@nomath\sc}

\begin{document}

\begin{flushright}
\small
LAPTH-036/20
\end{flushright}
\vskip0.5cm

\begin{center}
{\sffamily \bfseries \LARGE \boldmath
Composite Dark Matter and a horizontal symmetry}\\[0.8 cm]
{\large Alexandre Carvunis, Diego Guadagnoli, M\'eril Reboud, Peter Stangl} \\[0.5 cm]
\small
{\em Laboratoire d'Annecy-le-Vieux de Physique Th\'eorique UMR5108\,, Universit\'e de Savoie Mont-Blanc et CNRS, B.P.~110, F-74941, Annecy Cedex, France}
\end{center}

\medskip

\begin{abstract}
We present a model of composite Dark Matter (DM), in which a new QCD-like confining ``hypercolor'' sector generates naturally stable hyperbaryons as DM candidates and at the same time provides mass to new weakly coupled gauge bosons $H$ that serve as DM mediators, coupling the hyperbaryons to the Standard Model (SM) fermions.
By an appropriate choice of the $H$ gauge symmetry as a horizontal $SU(2)_h$ SM flavor symmetry, we show how the $H$ gauge bosons can be identified with the horizontal gauge bosons recently put forward as an explanation for discrepancies in rare $B$-meson decays.
We find that the mass scale of the $H$ gauge bosons suggested by the DM phenomenology intriguingly agrees with the one needed to explain the rare $B$-decay discrepancies.
\end{abstract}

\vspace{0.8cm}

\renewcommand{\thefootnote}{\arabic{footnote}}
\setcounter{footnote}{0}

\noindent

\section{Introduction}

New massive particles that only weakly interact with the Standard Model (SM) particles, the so-called weakly interacting massive particles (WIMPs), are among the most promising explanations of Dark Matter (DM)~\cite{Bertone:2004pz,Roszkowski:2017nbc}.
One main challenge in constructing viable WIMP models is the requirement that the DM candidate has to be stable on time scales of the order of the age of the Universe.
Quite generically, such stability requires the presence of an approximately unbroken global symmetry, one famous example being $R$-parity in supersymmetric DM models.
In the SM, the only global symmetry that provides stability to a heavy particle is the $U(1)_B$ baryon symmetry.
It protects the lightest baryonic bound state from decaying to leptons, thus providing the proton with a mean lifetime of more than $10^{29}$ years~\cite{Tanabashi:2018oca}.
An immediate question is then whether in an extension of the SM, a symmetry similar to the $U(1)_B$ baryon symmetry could provide stability to a DM candidate.
Such an extension can be given by a new QCD-like confining ``hypercolor'' (HC) sector, in which hyperquarks are charged under an accidental $U(1)_{\HB}$ hyperbaryon symmetry.
These hyperquarks form baryonic bound states, the hyperbaryons, which are rendered stable by carrying non-zero $U(1)_{\HB}$ charges and are thus potential DM candidates.\footnote{Several models with a composite DM made stable by a similar symmetry mechanism have been proposed. Those we are aware of include
\cite{Nussinov:1985xr,Chivukula:1989qb,Barr:1990ca,Barr:1991qn,Kaplan:1991ah,Gudnason:2006yj,Foadi:2008qv,Khlopov:2008ty,Ryttov:2008xe,Sannino:2009za,Bai:2010qg,Buckley:2012ky,Cline:2013zca,Antipin:2014qva,Antipin:2015xia,Mitridate:2017oky,Bai:2018dxf,Farrar:2018hac,Gross:2018ivp,Farrar:2020zeo}.
For a review, see \cite{Kribs:2016cew}.}

In addition to hyperbaryons, such a construction also leads to hypermeson bound states formed by a hyperquark and an antihyperquark.
Carrying zero $U(1)_{\HB}$ charge, the hypermesons are in general unstable.
Furthermore, without an explicit breaking of the hyperquarks' chiral symmetry,
the lightest hypermesons are massless Nambu-Goldstone bosons (NGBs).
The appearance of phenomenologically unacceptable massless scalars can be avoided in two ways.
Lagrangian terms that explicitly break the chiral symmetry can generate a potential for the NGBs, turning them into massive pseudo NGBs (pNGBs)~\cite{Weinberg:1972fn}.
An example are the pions in QCD, which are massive because the quark mass terms explicitly break the quarks' chiral symmetry.
The second possibility to avoid massless NGBs is the gauging of an appropriate subgroup of the chiral symmetry.
In this case, the corresponding gauge bosons become massive due to the Higgs mechanism and the would-be NGBs become the gauge bosons' longitudinal degrees of freedom.
Among these two possibilities, the latter one is particularly interesting in the context of a DM model since the gauging of a subgroup of the chiral symmetry implies that the corresponding gauge bosons, which we denote by $H$, couple to the hyperbaryon DM candidates.
If these $H$ gauge bosons also couple to some of the SM particles, they then not only solve the problem of massless NGBs, but they could also serve as mediators to a dark sector naturally endowed with a stability mechanism.

Interestingly, in this setup, the masses of the hyperbaryons $\mcX$ are related to those of the gauge bosons $H$ by
\be
\label{eq:mChi_mH_relation}
m_\chi \approx 10\, v_\HC \qquad \text{where} \qquad v_\HC = 2 m_H / g_H
\ee
with $g_H$ the $H$ gauge coupling and $v_\HC$ the NGB decay constant associated to the dynamical breaking of the chiral symmetry by a hyperquark condensate.
Unitarity arguments suggest that the mass of a thermal relic DM candidate cannot exceed 340 TeV~\cite{Griest:1989wd}.
Taking the hyperbaryons $\mcX$ as DM candidates, eq.~\eqref{eq:mChi_mH_relation} then implies
\begin{equation}\label{eq:mH_bound}
 v_{\HC} \lesssim 34\ \text{TeV}\,.
\end{equation}
This puts the $H$ gauge bosons in the mass range probed by flavor physics experiments.
Intriguingly, current data on rare $B$-meson decays shows deviations from the SM predictions in several measurements, most notably in the theoretically clean lepton flavor universality ratios $R_{K^{(*)}}$~\cite{Aaij:2017vbb,Aaij:2019wad,Abdesselam:2019wac,Abdesselam:2019lab}, which hint at new physics (NP) at a scale well compatible with eq.~\eqref{eq:mH_bound}~\cite{DiLuzio:2017chi}.
The discrepancies in rare $B$ decays can be explained by NP contributions to the $b\to s\ell\ell$ transition~\cite{Aebischer:2019mlg,Alguero:2019ptt,Ciuchini:2019usw,Datta:2019zca,Kowalska:2019ley,Arbey:2019duh}, which can be generated at tree level either by leptoquarks or SM-neutral bosons.
An immediate question is whether the $H$ gauge bosons can play the role of such tree-level mediators contributing to the $b\to s\ell\ell$ transition.
They could be leptoquarks only if the chiral symmetries of the hyperquarks contained at least the SM $SU(3)_C$ and $U(1)_Y$ groups as subgroups. However, in this case the hyperbaryons would in general not be SM-neutral DM candidates.\footnote{In a model featuring a leptoquark that explains the $B$-discrepancies and gets its mass from a QCD-like hypercolor sector, SM-neutral hyperbaryons could be formed e.g.\ from two types of hyperquarks transforming in different representations of the HC gauge group~\cite{Fuentes-Martin:2020bnh}.}
On the other hand, in the case of SM-neutral $H$ gauge bosons, the hyperbaryons are naturally SM-neutral, thus allowing them to be viable DM candidates.
A main challenge for models explaining the $b\to s\ell\ell$ discrepancies in terms of SM-neutral tree-level mediators are too large NP contributions to $B_s$-$\bar{B}_s$ mixing~\cite{DiLuzio:2019jyq}.
However, such large contributions can be avoided if the mediators are the gauge bosons of a gauged horizontal $SU(2)_h$ flavor symmetry~\cite{Guadagnoli:2018ojc}.
We are thus led to a well-motivated model with the following features:
\begin{itemize}
 \item A new QCD-like confining hypercolor sector generates baryonic bound states of hyperquarks, the hyperbaryons, which are SM-neutral and stable and are thus natural DM candidates.
 \item Massless NGBs are avoided by gauging an appropriate subgroup of the hyperquarks' chiral symmetry, resulting in massive gauge bosons $H$.
 \item The $H$ gauge bosons are coupled to SM fields by identifying the $H$ gauge group with the diagonal subgroup of part of the hyperquarks' chiral symmetry and a horizontal $SU(2)_h$ flavor symmetry of SM fermions.
 This allows the $H$ gauge bosons to address the discrepancies in rare $B$-decays as proposed in~\cite{Guadagnoli:2018ojc}.
 \item The $H$ gauge bosons couplings to hyperquarks and SM fermions as described above allows them to play the role of the DM mediators, opening the possibility of a fully calculable DM freeze-out scenario with the hyperbaryons as DM candidates.
\end{itemize}
The remaining parts of this paper analyzes the DM phenomenology of this construction.
The details of the model are discussed in Section~\ref{sec:model}.
In Section~\ref{sec:DD} we investigate possible bounds from DM direct detection.
Our prediction for the DM relic density and a discussion of the viable parameter space allowing for an explanation of the observed relic density are presented in Section~\ref{sec:Omega_DM}.
In Section~\ref{sec:Conclusions} we conclude.

\section{Model}\label{sec:model}

In order to illustrate the idea in our paper, let us start from a known example: two-flavor QCD.
In the absence of explicit mass terms, the left- and right-handed $u$ and $d$ quarks transform as doublets, $(u_L \,,\, d_L)^T$ and $(u_R \,,\, d_R)^T$, under the global symmetry groups $SU(2)_L$ and $SU(2)_R$, respectively.
The global chiral symmetry group is therefore $\mathcal G_\chi = SU(2)_L \times SU(2)_R$.
This symmetry is
spontaneously broken to $\mathcal H_V = SU(2)_V$ by the quark condensate $\< \bar u_L u_R + \bar d_L d_R + {\rm H.c.} \> \neq 0$ at the scale $f_\pi \simeq 93$ MeV.
The consequence are Goldstone bosons spanning the coset space $\mathcal G_\chi / \mathcal H_V$, the known pions.
As well known, the $SU(2)_L$ group is actually gauged, and this gauge symmetry is spontaneously broken {\em also} by the Higgs vev $v = 246$ GeV.
Because of the hierarchy $f_\pi \ll v$, the pions contribute negligibly to the longitudinal degrees of freedom of the $SU(2)_L$ gauge bosons.

Pursuing the analogy, one may introduce two flavors of new massless vector-like fermions, $\mathcal{F}_{2}$ and $\mathcal{F}_{3}$, charged under a new strong force, corresponding to the gauge group $\mathcal G_{\HC} \equiv SU(N)_{\HC}$ that will be denoted by \emph{hypercolor}.\footnote{Following existing literature, this naming denotes vector-like confinement, or a new chiral-symmetric QCD-like force, see in particular \cite{Kilic:2009mi,Pasechnik:2013bxa,Lebiedowicz:2013fta,Pasechnik:2014ida,Matsuzaki:2017bpp}.}
In order to be able to use known results from QCD for otherwise incalculable quantities (cf. Sec. \ref{sec:sigmaeff_v}), we fix $N=3$.
Just like in two-flavor QCD, these new fermions transform as doublets $\mathcal{F}_L=
(\mathcal{F}_{2\,L} \,,\, \mathcal{F}_{3\,L})^T
$
and
$\mathcal{F}_R=
(\mathcal{F}_{2\,R} \,,\, \mathcal{F}_{3\,R})^T
$
under the global symmetry groups $SU(2)^\mathcal{F}_L$ and $SU(2)^\mathcal{F}_R$, respectively.
The chiral condensate $\< \bar{\mathcal{F}}_{2\,L} \mathcal{F}_{2\,R} + \bar{\mathcal{F}}_{3\,L} \mathcal{F}_{3\,R} + {\rm H.c.}\> \propto v_{\HC}^3$ \footnote{Note that in QCD one has $\langle \bar{u}u + \bar{d}d\rangle = -2\,B\,f^2 \approx -16\pi\,f^3$.}
breaks the chiral symmetry group $G^{\mc F} \equiv SU(2)^\mathcal{F}_L \times SU(2)^\mathcal{F}_R$ to the diagonal $SU(2)^\mathcal{F}_V$.
Then, if either $SU(2)^\mathcal{F}_L$ or $SU(2)^\mathcal{F}_R$ is gauged, this gauge symmetry, to be denoted by $SU(2)_h$, is also spontaneously broken by the chiral condensate.
As a consequence, the $SU(2)_h$ gauge bosons $H_\mu^{a}$ acquire mass, and the would-be Goldstone bosons of the spontaneously broken group supply the $H_\mu^a$ longitudinal polarizations.
One therefore ends up with a $H_\mu^a$ boson triplet, whose transverse polarizations are elementary, while their longitudinal ones are composite $\bar{\mathcal{F}} \mathcal{F}$ bound states.

Concretely, it is $SU(2)^\mathcal{F}_L$ that we choose to gauge, and identify with $SU(2)_h$, the horizontal gauge symmetry introduced in Ref.~\cite{Guadagnoli:2018ojc}.
There, it is assumed that 2$^{\rm nd}$- and 3$^{\rm rd}$-generation {\em left-handed} SM fermions transform as doublets under $SU(2)_h$
\be
\label{eq:hsym_SM_doublets}
Q^\prime \equiv (q^\prime_2, q^\prime_3)^T~,~~~~ L^\prime \equiv (l^\prime_2, l^\prime_3)^T~,
\ee
with $q^\prime_{2,3}$ and $l^\prime_{2,3}$ denoting the SM quark and lepton doublets of generation index $2,3$.\footnote{As we will discuss below, the prime denotes SM fields before EW-scale flavor mixing and also before hypercolor-scale mixing with vector-like new fermions.} Such construction was advocated in Ref.~\cite{Guadagnoli:2018ojc} to account for discrepancies in $b \to s \ell \ell$.
More quantitatively, the $SU(2)_h \times SU(2)^\mathcal{F}_R$ symmetry-breaking scale $v_{\HC}$ can be related to the effective scale pointed to by the SM $b \to s \ell \ell$ Hamiltonian $\mathcal H_{\rm SM}^{b \to s}$. Such scale bounds the ratio between the mass of the horizontal bosons $m_H$, and the coupling $g_H$ of the horizontal symmetry \cite{Guadagnoli:2018ojc}.
Using $m_H = g_H v_{\HC} / 2$, this translates into the indicative range
\be
\label{eq:vHC_range}
v_{\HC} \in [10, 30]~{\rm TeV}~,
\ee
which is nicely compatible with the bound in eq. (\ref{eq:mH_bound}).

Due to the fact that LH SM fermions are charged under $SU(2)_h$ while RH SM fermions are not, the usual Higgs Yukawa couplings are forbidden by gauge invariance and can only arise after the spontaneous breaking of $SU(2)_h$.
Two options to accomplish this are:
\begin{itemize}
 \item The SM Higgs is part of a larger scalar multiplet that transforms non-trivially under $SU(2)_h$ in such a way that it can have gauge-invariant Yukawa couplings with LH and RH SM fermions.
 This option would lead to an enlarged Higgs sector.
 \item The SM fermions mix with new fermions that are vector-like (VL) under the SM gauge group and transform non-trivially under $SU(2)_h$ in such a way that gauge-invariant Higgs Yukawa couplings between the new VL fermions and SM fermions are possible.
 This option allows for a minimal Higgs sector but requires an enlarged fermion sector.
\end{itemize}
In the following, we focus on the latter option since an enlarged fermion sector is required also for another reason.
In a model containing only the SM fermions and the hyperquarks $\mathcal{F}$, there would be an $SU(2)_h$ gauge anomaly in case of an $SU(N)_{\HC}$ group with odd $N$ \cite{Witten:1982fp}, in particular for our assumption $N=3$.
In this case, the $SU(2)_h$ gauge anomaly has to be canceled by an enlarged fermion sector containing an odd number of $SU(2)_h$ doublets.
We therefore introduce new RH fermions $\Psi_R$ with the same SM quantum numbers as one generation of RH SM fermions but transforming as doublets under $SU(2)_h$.
In addition, we introduce two sets of new LH fermions $\Psi_L^i$ with the same SM quantum numbers as $\Psi_R$ but transforming trivially under  $SU(2)_h$.
Furthermore, we assume the presence of an `extended' hypercolor (EHC) sector that generates at least one of the four-fermion operators\footnote{Here we use the denomination `extended' similarly as in extended technicolor \cite{Dimopoulos:1979es,Eichten:1979ah}, namely to denote dynamics designed to account for fermion masses.}
\begin{equation}\label{eq:EHC}
 \frac{c_{ij}}{\Lambda_{\EHC}^2}
 (\bar{\Psi}_L^i \Psi_R)
 (\bar{\mathcal{F}}_L \mathcal{F}_R^j)
 +h.c.\,,
 \quad\quad
 \frac{\tilde{c}_{ij}}{\Lambda_{\EHC}^2}
 (\bar{\Psi}_L^i \Psi_R)
 (\bar{\mathcal{F}}_R^j \mathcal{F}_L)
 +h.c.\,,
\end{equation}
which both generate a vector-like mass term for $\Psi$ once the hyperquarks condense.
All fermions in the model and their quantum numbers are collected in table~\ref{tab:fermions}.
The possible gauge invariant and renormalizable couplings between SM fermions, new fermions $\Psi$, and the Higgs field $\varphi$ are given by
\begin{equation}
\begin{aligned}
\label{eq:LPsiQ}
\mathscr L
\supset
&- \Delta_u^{ik} \ \bar{\Psi}^{\prime\,u\,i}_L\, u^{\prime\,k}_R
 - \Delta_d^{ik} \ \bar{\Psi}^{\prime\,d\,i}_L\, d^{\prime\,k}_R
 - \Delta_e^{ik} \ \bar{\Psi}^{\prime\,e\,i}_L\, e^{\prime\,k}_R
\\
&- y_u^k \ \bar{q}^{\prime\,1}_L \, \tilde \varphi \, u^{\prime\,k}_R
 - y_d^k \ \bar{q}^{\prime\,1}_L \, \varphi \, d^{\prime\,k}_R
 - y_e^k \ \bar{l}^{\prime\,1}_L \, \varphi \, e^{\prime\,k}_R
 \\
&- y_U \ \bar{Q}^\prime_L \, \tilde \varphi \, \Psi_R^{\prime\,u}
 - y_D \ \bar{Q}^\prime_L \, \varphi \, \Psi_R^{\prime\,d}
 - y_E \ \bar{L}^\prime_L \, \varphi \, \Psi_R^{\prime\,e}
 \,,
\end{aligned}
\end{equation}
where $i\in\{2,3\}$, $k\in\{1,2,3\}$.
\begin{table}[t]
\begin{center}
\def\arraystretch{1.4}
\begin{tabular}{|c|c|c|c|c|c|}
\hline
Field & $SU(3)_{\HC}$ & $SU(2)_{h}$ & $SU(3)_c$ & $SU(2)_L$ & $U(1)_Y$ \\
\hline
\hline
$Q_L^{\prime}$&${\bf 1}$ &${\bf 2}$ & ${\bf 3}$ & ${\bf 2}$ & $+1/6$ \\
$L_L^{\prime}$&${\bf 1}$ &${\bf 2}$ & ${\bf 1}$ & ${\bf 2}$ & $-1/2$ \\
$q_L^{\prime\,1}$&${\bf 1}$ &${\bf 1}$ & ${\bf 3}$ & ${\bf 2}$ & $+1/6$ \\
$\ell_L^{\prime\,1}$&${\bf 1}$ &${\bf 1}$ & ${\bf 1}$ & ${\bf 2}$ & $-1/2$ \\
$u_R^{\prime\,1,2,3}$&${\bf 1}$ &${\bf 1}$ & ${\bf 3}$ & ${\bf 1}$ & $+2/3$ \\
$d_R^{\prime\,1,2,3}$&${\bf 1}$ &${\bf 1}$ & ${\bf 3}$ & ${\bf 1}$ & $-1/3$ \\
$e_R^{\prime\,1,2,3}$&${\bf 1}$ &${\bf 1}$ & ${\bf 1}$ & ${\bf 1}$ & $-1$ \\
\hline
\hline
$\Psi_R^{\prime\,u}$&${\bf 1}$ &${\bf 2}$ & ${\bf 3}$ & ${\bf 1}$ & $+2/3$ \\
$\Psi_R^{\prime\,d}$&${\bf 1}$ &${\bf 2}$ & ${\bf 3}$ & ${\bf 1}$ & $-1/3$ \\
$\Psi_R^{\prime\,e}$&${\bf 1}$ &${\bf 2}$ & ${\bf 1}$ & ${\bf 1}$ & $-1$ \\
$\Psi_L^{\prime\,u\,2,3}$&${\bf 1}$ &${\bf 1}$ & ${\bf 3}$ & ${\bf 1}$ & $+2/3$ \\
$\Psi_L^{\prime\,d\,2,3}$&${\bf 1}$ &${\bf 1}$ & ${\bf 3}$ & ${\bf 1}$ & $-1/3$ \\
$\Psi_L^{\prime\,e\,2,3}$&${\bf 1}$ &${\bf 1}$ & ${\bf 1}$ & ${\bf 1}$ & $-1$ \\
\hline
\hline
$\mathcal{F}_L$&${\bf 3}$ &${\bf 2}$ & ${\bf 1}$ & ${\bf 1}$ & $0$ \\
$\mathcal{F}_R^{2,3}$&${\bf 3}$ &${\bf 1}$ & ${\bf 1}$ & ${\bf 1}$ & $0$ \\
\hline
\end{tabular}
\end{center}
\caption{Quantum numbers of SM-like fermions (first block), new vector-like fermions $\Psi$ (second block), and hyperquarks $\mathcal{F}$ (third block).
}
\label{tab:fermions}
\end{table}

By way of SM- and $\Psi$-fermion field redefinitions in flavor space, the above mixing Lagrangian leads to the ordinary SM Yukawa terms. The detailed derivation is presented in the Appendix. The DM phenomenology to follow depends on the Yukawa sector only through the matrices leading from the primed basis to the mass eigenbasis for quarks. In fact, quarks only communicate with the DM sector through the exchange of horizontal bosons, i.e. through the interaction introduced explicitly in eq. (\ref{eq:L_ud_H}).

\subsection{NGBs and baryon Lagrangian: the DM sector}\label{sec:NGB}

The $SU(3)_{\HC}$-charged $\mathcal F_{L,R}$ fermions will form bound states.
Interestingly for the DM phenomenology to follow, and as discussed in the Introduction, among these states, the lightest baryonic ones will be stable because of the accidentally conserved hyperbaryon quantum number.
Having fixed the dimension $N$ of the hypercolor space, $N = 3$, the multiplicity of states is also fixed. It is easy to see that one obtains a quartet of spin-3/2 states, plus a doublet of spin-1/2 states, the latter being the lightest states.\footnote{Similarly as in Gell-Mann's `eightfold way', such multiplets are obtained by first taking the fully symmetric irrep under spin $\times$ flavor, in our case $SU(2 \times 2)$, which yields a ${\bf 20}$, and then decomposing the latter into a direct sum of irreps under $SU(2) \otimes SU(2)$, which yields $({\bf 2}, {\bf 2}) + ({\bf 4}, {\bf 4})$.} In analogy with the proton and the neutron of regular isospin, these two states will be
\be
\label{eq:chipn}
\chi_{p} \sim \mathcal F_2 \mathcal F_2 \mathcal F_3~,~~~~
\chi_{n} \sim \mathcal F_2 \mathcal F_3 \mathcal F_3~,
\ee
where $2,3$ label the flavor index.
Scaling up the proton mass $m_p$ (see also discussion in Sec. \ref{sec:sigmaeff_v}), we assume $m_\chi = m_p v_{\HC} / f_\pi$, with $f_\pi = 93$ MeV.

Let us now turn to the discussion of mesonic states.
$\bar{\mc{F}} \mc F$ states are analogous to QCD's pions in the sense that they are the result of chiral $SU(2)_L^\mathcal{F} \times SU(2)_R^\mathcal{F}$ symmetry breaking at the scale $v_\HC$, while the pions are the result of $SU(2)_L \times SU(2)_R$ breaking at the scale $f_\pi$.
Furthermore, the $SU(2)_L$ appearing in two-flavor QCD and the $SU(2)_L^\mathcal{F}$ appearing in the HC sector are both gauged.
However, there is also an important difference between the QCD pions and the hyperpions.
The $SU(2)_L$ gauge symmetry is actually also broken by the Higgs vev at the scale $v=246$~GeV.
Both symmetry breakings generate NGBs, resulting in three Higgs NGBs and three pion NGBs.
A linear combination of the Higgs NGBs and the pion NGBs forms the longitudinal polarizations of the $SU(2)_L$ gauge bosons, while the orthogonal linear combinations constitute the physical pions.
The latter become massive pNGBs due to the quark masses that are generated by the Higgs vev.
Since the scales $v$ and $f_\pi$ are vastly different and the mixing angle $\alpha$ between Higgs NGBs and pion NGBs is given by the tiny $\tan \alpha=f_\pi/v\approx4\times10^{-4}$, the longitudinal polarizations of the $SU(2)_L$ gauge bosons are essentially the Higgs NGBs and the physical pions are essentially the pion NGBs.
The situation is different in the HC sector, where the only symmetry breaking of the gauged $SU(2)_L^\mathcal{F}=SU(2)_h$ is the chiral symmetry breaking.
Consequently, there are only the three hyperpion NGBs and they become exactly the longitudinal polarizations of the $SU(2)_h$ gauge bosons.

We next discuss the Nambu-Goldstone boson (NGB) and baryon Lagrangian in detail. Our aim is to determine the $\chi_{p,n}$ interactions with SM matter, that are necessary for the DM phenomenology to follow.
Apart from the differences mentioned above, the general structure of the Lagrangian that describes the interactions between the $\chi_{p,n}$ baryons, the $\bar {\mc F} \mc F$ mesons and the $SU(2)_h$ bosons is largely analogous to the structure of the pion-nucleon Lagrangian~\cite{Georgi:1985kw,Gasser:1987rb}.
We will follow the line of argument and normalization conventions in Ref.~\cite{Scherer:2009bt}.

Given our chiral group $G^{\mc F} \equiv SU(2)_L^{\mc F} \times SU(2)_R^{\mc F}$, one introduces the field
\begin{equation}
\Omega = \exp\left(i \, \frac{2 \, \Pi}{v_{\HC}}\right)\,,
\end{equation}
with $\Pi$ the `hyperpion' field $\Pi = \Pi^a\,T^a\,,$ and $T^a=\tau^a/2$ with the Pauli matrices $\tau^a$.
Further denoting the elements of the global $SU(2)_L^{\mc F}$ and $SU(2)_R^{\mc F}$ symmetries as $L$ and $R$,
one has that $\Omega$ transforms as
\be
\Omega \to \Omega^\prime = R \, \Omega \, L^\dagger~.
\ee
The NGB Lagrangian at leading order in an expansion in derivatives is given by
\begin{equation}
 \mathcal{L}_{\NGB} = \frac{v_{\HC}^2}{4}\,{\rm tr}\!\left[
 (D_\mu\,\Omega)\,(D^\mu\,\Omega)^\dagger
 \right]\,,
\end{equation}
where the covariant derivative reads
\begin{equation}
 D_\mu\,\Omega =
 \partial_\mu\,\Omega
 - i\, r_\mu\, \Omega
 + i\, \Omega\, l_\mu\,.
\end{equation}
The currents $l_\mu$ and $r_\mu$ are in our case given by
\begin{equation}\label{eq:gauge_currents}
l_\mu = - g_H\, H^a_\mu\,T^a\,,
\qquad
r_\mu = 0\,.
\end{equation}
Consequently, the NGB Lagrangian can be written as
\begin{equation}\label{eq:L_NGB_3_terms}
 \mathcal{L}_{\NGB}
 =
 \frac{v_{\HC}^2}{4}\,{\rm tr}\!
 \left[
 \partial_\mu\,\Omega\,\partial^\mu\,\Omega^\dagger
 \right]
 +\frac{1}{2}\,m_H^2\,H_\mu^a\,H^{a\mu}
 -i\,m_H\,v_{\HC}\,{\rm tr}\!\left[H_\mu\,(\partial^\mu\,\Omega^\dagger)\,\Omega \right]\,,
\end{equation}
where the first term is the pure NGB Lagrangian, the second terms yields a mass for the vector bosons with $m_H = \frac{1}{2}\,g_H\,v_{\HC}$, and the third term leads to interactions between a vector boson and the NGBs.
Using the definition of $\Omega$, one can expand
\begin{equation}
 (\partial^\mu\,\Omega^\dagger)\,\Omega
 =
 \frac{-2\,i}{v_{\HC}}\partial^\mu\,\Pi
 +
 \frac{1}{2!}
 \left(\frac{-2\,i}{v_{\HC}}\right)^2
 \left[
 \Pi
 ,
 \partial^\mu\,\Pi
 \right]
 +
 \frac{1}{3!}
 \left(\frac{-2\,i}{v_{\HC}}\right)^3
 \left[
 \Pi
 ,
 \left[
 \Pi
 ,
 \partial^\mu\,\Pi
 \right]
 \right]
 +...
\end{equation}
such that the third term in eq.~\eqref{eq:L_NGB_3_terms} becomes
\begin{equation}
 \mathcal{L}_{\NGB}\supset
 -m_H\,H_\mu^a\,\partial^\mu\Pi^a
 -\frac{g_H}{2}\,\varepsilon^{abc}\,H_\mu^a\,\Pi^b\,\partial^\mu\Pi^c
 +...\ .
\end{equation}
We observe that this leads to an inconvenient term that is linear in both $H_\mu^a$ and $\Pi^a$ and corresponds to a mixing between gauge bosons and NGBs.
However, this term can be removed by a gauge transformation.
To this end, we fix the gauge by adding the gauge-fixing Lagrangian
\begin{equation}
 \mathcal{L}_{GF}
 =
 -\frac{1}{\xi_H}\,{\rm tr}\!\left[G_H\,G_H\right]\,,
 \quad
 \text{where}
 \quad
 G_H = \partial^\mu H_\mu+\xi_H\,m_H\,\Pi\,,
\end{equation}
which results in
\begin{equation}
 \mathcal{L}_{GF}
 =
 -\frac{1}{2\,\xi_H}\,\partial^\mu H_\mu^a\,\partial^\mu H_\mu^a
 -\frac{1}{2}\,\xi_H\,m_H^2\,\Pi^a\,\Pi^a
 +m_H\,H_\mu^a\,\partial^\mu\Pi^a\,.
\end{equation}
The first term yields the gauge-fixing condition for $H_\mu^a$, the second term is a $\xi_H$-dependent mass for the $\Pi^a$ fields, and the third term exactly cancels the unwanted mixing term in $\mathcal{L}_{\NGB}$.

We now turn to the matter field $\mcX \equiv (\chi_{p}, \chi_{n})$. The latter transforms linearly as the fundamental irrep under the isospin subgroup $SU(2)^{\mc F}_V$ with elements $V$.
We can therefore write
\be
\mcX \to V \mcX~.
\ee
Having defined our dynamical fields $\Omega$ and $\mcX$ and their chiral-symmetry transformation properties, we can now introduce the leading effective `hyperbaryon' Lagrangian. First, we introduce the $\mcX$ covariant derivative
\be
\label{eq:DX}
i D_\mu \mcX \equiv (i\partial_\mu + i\Gamma_\mu + v_\mu^{(s)}) \mcX
\ee
such that $D_\mu \mcX$ transforms like $\mcX$.
Here
\be
\label{eq:Gamma_mu}
i\Gamma_\mu \equiv \frac{i}{2}
\left[
\omega^\dagger (\partial_\mu - i r_\mu) \omega + \omega (\partial_\mu - i l_\mu) \omega^\dagger
\right]~,
\ee
where the external gauge currents $r_\mu$ and $l_\mu$ are given in eq.~\eqref{eq:gauge_currents}, and in our case $v_\mu^{(s)}=0$.
The field $\omega$ is defined such that $\omega^2 =\Omega$ and it transforms as
\be
\omega \to \omega^\prime = R ~ \omega ~ K^{-1}(L, R, \omega)~,
\ee
with $K$ an $SU(2)$-valued function of its arguments, in particular $\omega(x)$-dependent. In the case of an {\rm isospin} transformation $R = L = V$ one has $K(V, V, \omega) = V$, namely the $\omega$ dependence drops.
The most general Lagrangian describing pion-nucleon interactions with the smallest number of derivatives takes the form
\be \label{eq:LpiN}
    \mathcal{L}_{\Pi \mcX} = \bar{\mcX} \left( i \slashed{D} - m_{\chi} + g_A\,\gamma^{\mu}\gamma_{5} \frac{u_{\mu}}{2} \right) \mcX
\ee
where $u_\mu$ is the so-called vielbein defined as
\be \label{eq:u_mu}
\frac{u_{\mu}}{2} \equiv \frac{i}{2} \left[
  \omega^\dagger (\partial_\mu - i r_\mu) \omega - \omega (\partial_\mu - i l_\mu) \omega^\dagger
  \right]
  ~.
\ee

The above Lagrangian generates interactions between $\chi$ and horizontal gauge bosons $H^a$ as well as $\bar{\mc{F}} \mc F$ mesons. Standard manipulations of the Lagrangian in eq.~(\ref{eq:LpiN}) lead to the following
interaction terms including one or two boson fields
\begin{equation}\label{eq:LpiN_explicit}
\begin{aligned}
\mathcal{L}_{\Pi\mcX}
\supset
\overline{\mcX} \Bigg[
&
- g_H\,\gamma^\mu\,H_\mu^a\left(T^a+\frac{1}{v_{\HC}}\,\varepsilon^{abc}\,\Pi^b\,T^c\right)\left(\frac{1 - g_A\gamma_5}{2}\right)
\\
&
+i\,g_H\,g_A\,\frac{m_\chi}{m_H}\,\gamma_5\,\Pi
+\frac{1}{2\,v_{\HC}^2}\,\gamma^\mu\,\varepsilon^{abc}\,(\partial_\mu\,\Pi^a)\,\Pi^b\,T^c
\Bigg] \mcX~,
\end{aligned}
\end{equation}
where the parameters $m_\chi$ and $g_A$ are the chiral limits of the $\mcX$-multiplet mass and of the axial vector coupling of the multiplet to the hyperpions.

From eq. (\ref{eq:LpiN_explicit}) we can read off all the $\chi \chi$ interactions that are necessary for the phenomenology to follow, in particular horizontal-boson exchange -- relevant to DM direct detection -- as well as one- or two-boson exchange -- in principle relevant to the DM relic abundance. As concerns the latter case -- discussed in detail in Sec. \ref{sec:sigmaeff_v} -- the problem of estimating the relevant cross section resembles very closely the determination of baryon-antibaryon annihilation close to threshold, for which we have abundant data. These data actually suggest that the $\chi \chi$-annihilation amplitude that one can estimate perturbatively from eq. (\ref{eq:LpiN_explicit}) represents a sub-dominant contribution with respect to other contributions that cannot be captured with the $\mc L_{\Pi \mcX}$ Lagrangian, and that will be estimated directly from data.

\section{Dark-Matter Direct Detection}\label{sec:DD}

Both the DM multiplet and quarks interact with horizontal bosons. Therefore DM can scatter onto nuclei via the tree exchange of $SU(2)_h$ bosons, and DM direct-detection (DD) data may in principle offer a separate probe of the $v_{\HC}$ scale.

The DM-$H_\mu^a$ interaction relevant for such process may be read off from eq. (\ref{eq:LpiN_explicit}). As concerns (light) quark-$H_\mu^a$ interactions, also relevant for such process, the discussion in App.~\ref{sec:SM_VL_fermions} yields the following Lagrangian terms
\bea
\label{eq:L_ud_H}
\mc L_{qH} &=& - g_H \Bigl(
\bar{u}_L \, U^\dagger_L \,  \slashed{H} \, U_L \, u_L +
\bar{d}_L \, D^\dagger_L \, \slashed{H} \, D_L \, d_L \nn \\
&+& \bar{u}_R \, U^\dagger_R U_R^{\Psi^u u \, \dagger} \, \slashed{H} \, U_R^{\Psi^u u} U_R \, u_R +
\bar{d}_R \, D^\dagger_R U_R^{\Psi^d d \, \dagger} \, \slashed{H} \, U_R^{\Psi^d d} D_R \, d_R
\Bigl)~,
\eea
where $u, d$ denote up- and down-type quark fields in the mass eigenbasis. For ease of notation, we will henceforth use the redefinitions
\bea
\label{eq:UD_redefs}
&U_{L} \rightarrow \mc U_{L}~,&D_{L} \rightarrow \mc D_{L}~,\nn \\
&U_{R}^{\Psi^u u} U_R \rightarrow \mc U_{R}~,&U_{R}^{\Psi^d d} D_R \rightarrow \mc D_{R}~.
\eea

Since the momentum flowing through the $H$-boson propagator is way below its mass, one can write the following local interaction Lagrangian
\bea
\label{eq:L_eff}
&&\mc L_{\rm eff} ~=~ \nn \\
&&- \frac{g_H^2}{m_H^2} \left[ \bar \mcX \gamma^\mu T^a \left( \frac{1 - g_A \gamma_5}{2} \right) \mcX \right] \left[ \left( \bar u_L \, \mc U_L^\dagger \gamma_\mu \mc T^a \mc U_L \, u_L  + \{ L \rightarrow R \}\right) + \{ u \rightarrow d~, \mc U \rightarrow \mc D \} \right]~,\nn \\
\eea
where in the first squared bracket $T^a$ denote the $SU(2)_h$ generators, whereas in the second squared bracket $\mc T^a$ are $3 \times 3$ matrices defined by
\be
\mc T^a \equiv
\left(
\begin{array}{cc}
0 & \\
 & T^a
\end{array}
\right)~.
\ee

The quark bilinears in eq. (\ref{eq:L_eff}) are to be evaluated between external nucleons. Under the assumption (whose validity we will discuss below) that both initial- and final-state nucleons are at rest, the required matrix elements may be parameterized as follows
\bea
\label{eq:NNqq}
\< N(p') | \bar q \gamma^\mu q | N(p) \>|_{\vec{p} = \vec{p}^{\,\prime} = 0} &=& F_1^{q/N}(0) \,\,  \bar u_{N}(p') \gamma^\mu u_{N}(p)~,\nn \\
\< N(p') | \bar q \gamma^\mu \gamma_5 q | N(p) \>|_{\vec{p} = \vec{p}^{\,\prime} = 0} &=& F_A^{q/N}(0) \,\, \bar u_{N}(p') \gamma^\mu \gamma_5 u_{N}(p)~.
\eea
The forward $F_{1,A}$ form factors can be parameterized as customary (see e.g. \cite{Bishara:2017nnn})
\bea
F_1^{q/N}(0) = n_q^N~,\nn \\
F_A^{q/N}(0) = \Delta q^N~,
\eea
where $n_q^N$ counts the number of {\em valence} quarks $q$ within nucleon $N$ (e.g. $n_s^p = 0$). Besides we use
\be
\Delta u^p = 0.90~,~~~~\Delta d^p = -0.38~,~~~~\Delta s^p = -0.03~,
\ee
with $\Delta q^n$ obtained by isospin exchange.

The DM - $N$ (with $N = p,n$) elastic cross section can then be written as
\be
\label{eq:sig^N}
\sigma^N ~=~ \frac{M_N^2}{\pi v_{\HC}^4} \left( \mc V_N + 3 g_A^2 \mc A_N \right)~,
\ee
where the two terms on the r.h.s. are usually denoted as spin-independent (SI) and respectively spin-dependent (SD) in the literature. The constants $\mc V_N, \mc A_N$ are defined as
\bea
&& \mc V_N ~\equiv~ \frac{1}{2} \sum_a |X_{VN}^a|^2~, \nn \\
&& \mc A_N ~\equiv~ \frac{1}{2} \sum_a |X_{AN}^a|^2~,
\eea
with
\bea
&&X_{VN}^a ~=~ n_u^N (\overline U_R^a + \overline U_L^a) + n_d^N (\overline D_R^a + \overline D_L^a)~, \nn \\
[0.2cm]
&&X_{AN}^a ~=~ \Delta u^N (\overline U_R^a - \overline U_L^a) + \Delta d^N (\overline D_R^a - \overline D_L^a) + \Delta s^N (\overline S_R^a - \overline S_L^a) ~,
\eea
and
\be
\label{eq:UDbar}
\overline U^a_{L,R} ~\equiv~ \left( \mc U_{L,R}^\dagger \mc T^a \mc U_{L,R} \right)_{11}~,~~~
\overline D^a_{L,R} ~\equiv~ \left( \mc D_{L,R}^\dagger \mc T^a \mc D_{L,R} \right)_{11}~,~~~
\overline S^a_{L,R} ~\equiv~ \left( \mc D_{L,R}^\dagger \mc T^a \mc D_{L,R} \right)_{22}~.
\ee
A numerical survey of eq. (\ref{eq:sig^N}) yields
\be
\label{eq:sig^N_approx}
\sigma^N \simeq 1.09 \left( \frac{10~{\rm TeV}}{v_{\HC}} \right)^4 10^{-44} ~{\rm cm}^2 \times [0, 3]~,
\ee
where the last factor on the r.h.s. provides the range of values spanned by the $\mc V_N + 3 g_A^2 \mc A_N$ factor in eq. (\ref{eq:sig^N}) as the $\mc U_{L,R}$ and $\mc D_{L,R}$ matrices are scanned over with the constraint $\mc U_L^\dagger \mc D_L = V_{\rm CKM}$ (and no other constraint, in particular from flavor observables).\footnote{In this scan we took $g_A = 1.27$, as in QCD. In doing so, we assume that QCD's $g_A$ at scales well below QCD's confinement scale should be similar to HC's $g_A$ at scales well below HC's confinement scale. This should be the case if $N_{\HC} = N_c$, if the non-perturbative dynamics is in both cases dominated by the two lightest (hyper)quarks, and if one can neglect (hyper)quark masses. We also note that $\mc A_N \ll \mc V_N$, hence $g_A$ does not need to be fixed very precisely.}
We remark that, even for $\mc V_N + 3 g_A^2 \mc A_N$ as large as it gets, $\sigma^N$ remains safely below (a naive high-mass extrapolation of) the Xenon1T bound \cite{Aprile:2018dbl}, keeping in mind that for $v_{\HC} \ge 10$ TeV, one has $M_{\chi} \ge 100$ TeV.

As well known, the analytic approach leading to eq. (\ref{eq:sig^N}) has a number of limitations, that have been discussed in Refs. \cite{Fan:2010gt,Fitzpatrick:2012ix,Fitzpatrick:2012ib,Anand:2013yka,Cirigliano:2012pq,DelNobile:2013sia,Barello:2014uda,Hill:2014yxa,Hoferichter:2015ipa,Catena:2014uqa,Hill:2013hoa,Hill:2011be,Hoferichter:2016nvd,Kurylov:2003ra,Pospelov:2000bq,Bagnasco:1993st,Bishara:2016hek,Bishara:2017pfq}. Among them is the necessity to include renormalization-group running effects from the scale of the effective interaction between the DM and the quarks, down to the scale of the recoil energy that is measured in DM direct-detection experiments. This effect may be estimated by the {\tt DirectDM} \cite{Bishara:2016hek,Bishara:2017pfq,Bishara:2017nnn} code. Another potentially important effect is the departure of eqs. (\ref{eq:NNqq}) from the static limit, whereby one introduces form factors depending on $q^2$, with $q = p^\prime - p$. This effect may be quantified with the {\tt DMFormFactor} \cite{Fitzpatrick:2012ix,Fitzpatrick:2012ib,Anand:2013yka} code.

In our context, where the analytic procedure yields a direct-detection signal safely below experimental limits, the above effects would be relevant only if they enhanced the signal by orders of magnitude. Ref. \cite{Guadagnoli:2020tlx} found that actually the numerical estimate obtained with the {\tt DirectDM} plus {\tt DMFormFactor} codes is lower with respect to the analytic result, by a factor not exceeding 2 for $M_\chi \lesssim 1.5$ TeV and for the interaction strength considered in that work. We extended that analysis to our case where the interaction strength is of order $1/ v_{\HC}^2$, with $v_{\HC}$ in the range in eq. (\ref{eq:vHC_range}), and where $M_\chi = 10\, v_{\HC}$ (see below eq. (\ref{eq:chipn})). We find again that $\sigma^N_{\rm analytic} > \sigma^N_{\rm numerical}$, by a factor close to 4, i.e. even larger than in Ref. \cite{Guadagnoli:2020tlx}. This fact is not surprising, because even if the DM approaches the nucleus with a momentum of order $M_\chi v$, with $v \sim 10^{-3}$, the recoil energy does not increase indefinitely as $M_\chi$ increases, but instead it reaches the maximum value $M_{\mc N} (2 v)^2 / 2$, with $M_{\mc N}$ the mass of the nucleus $\mc N$.

\section{Dark Matter Relic Density}\label{sec:Omega_DM}

We wish to estimate the relic density of our DM multiplet $\mcX$ at present time, $\Omega_0 h^2$. The first step towards this end is the determination of $\Omega h^2$ at the `freeze-out' temperature $T_f$. In our case of a single, mass-degenerate, DM multiplet $\mcX$, $T_f$ is determined recursively from the `matching' relation \cite{Kolb:1990vq,Griest:1990kh}
\be
\label{eq:Tf}
x_f ~=~ \ln
\frac{0.038 g_{\rm eff} M_{\rm Pl} m_{\chi} \< \sigma_{\rm eff} v\>}{g_*^{1/2} x_f^{1/2}}~,
\ee
where $x_f = m_{\chi} / T_f$, $g^*$ is the number of effectively relativistic d.o.f. at $T_f$ and $g_{\rm eff}$ the number of internal d.o.f. of the $\mcX$, given in our case by
\be
\label{eq:g_eff}
g_{\rm eff} = 2 g_p~,
\ee
and $g_n = g_p$ is the number of internal d.o.f.\ of the $\chi_n$ or $\chi_p$ particles.\footnote{Remembering the definitions in eq. (\ref{eq:chipn}), $\chi_{p,n}$ although e.m.-neutral, cannot be Majorana fermions, hence $g_{p,n} = 4$.}

The main dynamical quantity, in eq.~(\ref{eq:Tf}) as well as in the other steps towards determining $\Omega_0 h^2$, is $\< \sigma_{\rm eff} v\>$.
In our case of mass-degenerate $\chi_i$ \cite{Griest:1990kh}
\be
\label{eq:sigmaeff}
\sigma_{\rm eff} \equiv \frac{1}{4} \sum_{i,j} \sigma_{ij}\,,
\ee
where $\sigma_{ij}$ is the annihilation cross section of $\bar{\chi}_i$ and $\chi_j$ and $i,j\in\{p,n\}$.

The second step towards obtaining $\Omega_0 h^2$ is to estimate the post-freeze-out annihilation efficiency $J$ \cite{Griest:1990kh,Jungman:1995df}
\be\label{eq:J}
J \equiv \int_{x_f}^\infty \frac{\< \sigma_{\rm eff} v\>}{x^2} dx~,
\ee
whence $\Omega_0 h^2$ can be estimated as\footnote{One can derive this relation by using $H(T) = \sqrt{8 \pi^3 g_* / 90} \,\, T^2 / M_{\rm Pl}$, $s = 2 \pi^2 g_* T^3 / 45$ and $\rho_c = 3 H_0^2 / (8 \pi G_N)$, with $H(T_0) \equiv H_0 = 100 h$ km$/$ (s Mpc), as customary.}
\be
\label{eq:Omega_DM}
\Omega_0 h^2 ~=~ \sqrt{\frac{45}{\pi}} \frac{s_0}{\rho_c} \frac{1}{g_*^{1/2} M_{\rm Pl} J}~\simeq~ \frac{1.07 \times 10^9 {\rm GeV}^{-1}}{g_*^{1/2} M_{\rm Pl}J}~.
\ee

\subsection{Calculation of $\< \sigma_{\rm eff} v\>$ and $J$} \label{sec:sigmaeff_v}

For composite baryon-like objects like the $\chi$, the dominant processes contributing to $\bar{\chi}\chi$ annihilation take place in the strongly coupled regime such that it is not straightforward to calculate the cross section from first principles.
In particular, as known from nucleon-antinucleon annihilation (for reviews, see e.g.~\cite{Green:1985vu,Dover:1992vj,Amsler:2019ytk,Richard:2019dic}), an important role is played by processes in which quarks and antiquarks inside the baryon and antibaryon are rearranged into mesons.
In this case, even without an actual annihilation of the quarks, baryon and antibaryon are ``annihilated''.
Other important processes are those in which some of the quarks and antiquarks actually annihilate via the strong interaction.
Given that these processes happen in the strongly coupled regime, other weakly coupled interactions like QED in the case of nucleon-antinucleon annihilation or the $SU(2)_h$ interactions in the case of $\bar{\chi}\chi$ annihilation play a subdominant role.

Since our strongly coupled sector resembles two-flavor QCD and the $\chi$ fields are very similar to proton and neutron in QCD, we estimate the $\bar{\chi}\chi$ annihilation cross section by scaling up experimental data on nucleon-antinucleon annihilation from the scale of the proton mass $m_p$ to the scale $m_\chi$ (a similar assumption was adopted in Ref. \cite{Antipin:2014qva}).
Before doing so, it is worth pointing out the expected differences between the QCD and HC cases that are not related to the overall scale. These differences -- that are sources of potential uncertainty -- include the following:
\begin{itemize}
 \item Nucleon-antinucleon annihilation data display a certain degree of isospin dependence, but we are unable to determine whether the latter is significant. In fact, in \cite{Kalogeropoulos:1980az} neutron-anti\-proton annihilation is roughly a factor $0.75$ smaller than proton-antiproton annihilation. While this might be an effect mostly at low $p_{\lab}$ due to electromagnetic attraction of $p$ and $\bar{p}$, we are not aware of any study that quantified such effect. On the other hand, according to \cite{Dover:1992vj} the differences between the proton-antiproton and neutron-anti\-proton annihilation cross section are not significant. As detailed below, we perform our DM-phenomenology calculations with several different fits, collected in table \ref{tab:sigma_ann}, and use the spread of the predictions obtained as an estimation of this error.

 \item In nucleon-antinucleon annihilation, approximately 5\% of the resulting mesons contain strange quarks (see e.g. \cite{Dover:1992vj} and references therein), which are not present in two-flavor QCD. Due to the absence of analogous annihilation channels in the case of $\bar{\chi}\chi$ annihilation, we should, in principle, lower the scaled-up cross section by 5\%, and add a 5\% one-sided uncertainty accounting for our naive treatment. In practice, since the uncertainties inherent in our DM-phenomenology calculations, notably that of the relic abundance, are not smaller than 10\% \cite{Griest:1990kh}, we neglect the above error.
 \item The masses of NGB final-state mesons are the consequence of Lagrangian terms that explicitly break the global diagonal $SU(2)_V^{(\mathcal{F})}$ (hyper)isospin symmetry.
 In QCD, isospin is explicitly broken by the quark masses, which leads to a ratio of the pion mass and decay constant $m_\pi/f_\pi\approx1.35$. Then, in nucleon-antinucleon annihilations such effect distorts the zero-meson-mass result by terms that are expectedly of order $(m_\pi / m_N)^2 \approx 2\%$.
 In our setup, hyperisospin is broken by the gauging of $SU(2)_h$ such that the hyperpions become the longitudinal polarizations of the horizontal gauge bosons. The ratio of their mass and the hyperpion decay constant $v_{\HC}$ is $m_H/v_{\HC}=g_H/2$.
 Therefore, for large $g_H$ the effect might be similar in size to the one in QCD, while for small $g_H$ the effect might be smaller than in QCD. We conclude that this effect may also be neglected.
 \item The masses of non-NGB final-state mesons like the vector mesons are non-zero also in the chiral limit. Furthermore, for (hyper)quark masses that are small with respect to the confinement scale, the masses of these mesons are essentially independent of the (hyper)quark masses (see e.g.~\cite{Yang:2014xsa} for a lattice study of the mass decomposition).
 In our setup, the masses and couplings of the vector hypermesons can slightly be affected by an expected mixing with the $SU(2)_h$ gauge bosons.
However, we expect that this mixing does not change the overall cross section significantly. We therefore overlook this source of systematic difference.
\end{itemize}
In the following, we will consider different fits to the nucleon-antinucleon annihilation cross section, and scale them up from the nucleon to the $m_\chi$ mass scale. As we argued, the spread across the predictions obtained from the different fits may provide a reasonable `envelope' for the theoretical uncertainty associated with the differences between QCD and our HC sector.
Nucleon-antinucleon annihilation has been studied by many different experiments.
Several groups provide fits to experimental data on the annihilation cross section $\sigma_{\ann}$ using the parameterization
\begin{equation}\label{eq:sigma_ann}
 \sigma_{\ann}=\frac{1}{m_p^2}\left(A+B\,\frac{m_p}{p_{\lab}}+C\,\frac{m_p^2}{p_{\lab}^2}\right)\,,
\end{equation}
where $p_{\lab}$ is the momentum of the antinucleon in the rest frame of the nucleon and $m_p$ is the proton mass.
Fit results are shown in table~\ref{tab:sigma_ann}.
In most fits, the coefficient $C$ is set to 0.
A non-zero $C$ can provide a slightly better fit at $p_{\lab}$ around 75~MeV~\cite{Astrua:2002zg}, corresponding to $v \sim 10^{-1}$. More generally, data used in the fits in table \ref{tab:sigma_ann} span the velocity range [0.05, 0.9], which includes the velocities $\sim 10^{-1}$ relevant to our freeze-out dynamics.

\begin{table}[t]
\begin{center}
\def\arraystretch{1.4}
\begin{tabular}{|c|c|c|c|c|c|c|}
\hline
Fit \# & Ref. & $N\bar{N}$ & $p_{\lab}$ [GeV] & $A$ & $B$ & $C$ \\
\hline
\hline
1 & \cite{Kalogeropoulos:1980az}
&$p\bar{p}$
&$[0.26,0.47]$
&$86$
&$84$
&$0$\\
2 & \cite{Bruckner:1987ve}
&$p\bar{p}$
&$[0.40,0.60]$
&$66.5 \pm 4.1$
&$77.1\pm2.2$
&$0$\\
3 & \cite{Brando:1985ft}
&$p\bar{p}$
&$[1.90,1.96]$
&$19.2 \pm 5.7$
&$98.1 \pm 3.1$
&$0$\\
\hline
\hline
4 & \cite{Kalogeropoulos:1980az,Bizzarri:1973sp}
&$n\bar{p}$
&$[0.26,0.47]$
&$63$
&$63$
&$0$\\
5 & \cite{Astrua:2002zg}
&$N\bar{n}$
&$[0.05,0.40]$
&$150.4\pm6.8$
&$48.0\pm2.2$
&$0$\\
6 & \cite{Astrua:2002zg}
&$N\bar{n}$
&$[0.05,0.40]$
&$199.9\pm10.6$
&$23.9\pm4.1$
&$2.5\pm0.4$
\\
\hline
\end{tabular}
\end{center}
\caption{Fit parameters describing experimental data on nucleon-antinucleon annihilation parameterized as $\sigma_{\ann}=\frac{1}{m_p^2}\left(A+B\,\frac{m_p}{p_{\lab}}+C\,\frac{m_p^2}{p_{\lab}^2}\right)$, where $p_{\lab}$ is the momentum of the antinucleon in the rest frame of the nucleon and $m_p$ is the proton mass. All dimensionful parameters in \cite{Kalogeropoulos:1980az,Bruckner:1987ve,Brando:1985ft,Astrua:2002zg} have been expressed in units of $m_p$ such that the coefficients $A$, $B$, and $C$ used here are dimensionless.
}
\label{tab:sigma_ann}
\end{table}

We assume that the $\bar{\chi}_i \chi_j$ annihilation cross section $\sigma_{ij}$ can be parameterized as in eq.~\eqref{eq:sigma_ann} with $m_p\to m_\chi$, and that it is hyper-isospin independent, i.e.\
$\sigma_{ij} = \sigma_{\ann}\big|_{m_p\to m_\chi}$,
which, using eq.~\eqref{eq:sigmaeff}, yields
\begin{equation}
 \sigma_{\rm eff} = \sigma_{\ann}\big|_{m_p\to m_\chi}\,.
\end{equation}
In order to compute the thermally averaged cross section $\< \sigma_{\rm eff} v\>$ and the post-freeze-out annihilation efficiency $J$, an expansion at low velocity and large $x=m_\chi/T$ is commonly employed~\cite{Srednicki:1988ce}. However, the low-velocity expansion does not converge for a cross section of the form as the one given in eq.~\eqref{eq:sigma_ann}.
Therefore, in the following, we perform the thermal averaging without such an expansion.
To this end, we consider the definition of $\< \sigma_{\rm eff} v\>$ in terms of the integral~\cite{Cannoni:2013bza}
\begin{equation}\label{eq:sig_eff_integral}
 \< \sigma_{\rm eff} v\> =
 \frac{4\,x}{K_2^2(x)}
 \int_1^\infty dy\, \sqrt{y}\,(y-1)\,K_1(2x\sqrt{y})\,\sigma_{\rm eff}(y)\,,
\end{equation}
where $K_\alpha$ is the modified Bessel function of the second kind of order $\alpha$, and we introduce the dimensionless variable $y=s/(4 \,m_\chi^2)$ with the Mandelstam variable $s$ denoting the square of the $\bar{\chi}\chi$ center-of-mass energy.
Changing variables in eq.~\eqref{eq:sigma_ann} using $p_{\lab}=2\,m_\chi\,\sqrt{y^2-y}$, the integral~\eqref{eq:sig_eff_integral} can be expressed as
\begin{equation}
\label{eq:sigmaeffv_ABCD}
 \< \sigma_{\rm eff} v\> =
 \frac{A\,f_0(x)+B\,f_{-1}(x)+C\,f_{-2}(x)}{m_\chi^2}\,
\end{equation}
where the functions $f_k(x)$ can be given in terms of Meijer's $G$ functions and the subscript $k$ corresponds to the power of $p_{\lab}$ in eq.~\eqref{eq:sigma_ann}.
Since $\< \sigma_{\rm eff} v\>$ is evaluated only for $x\geq x_f\approx30$, it is convenient to expand $f_k(x)$ in powers of $1/x$, which yields
\begin{equation}
 f_k(x) = \frac{2^{2+k}\,\Gamma(2+k/2)}{\sqrt{\pi}} \,x^{-(1+k)/2}\left(1+\sum_{n=1}^\infty \frac{c_n(k)}{x^n}\right)\,,
\end{equation}
where the first three coefficients $c_n(k)$ are given by\footnote{%
We verified that this series converges so quickly that terms beyond $c_1$ provide corrections way below 1\%.}
\begin{equation}
\begin{aligned}
 c_1(k) &= -\frac{25}{16}+\frac{7}{4}\,k+\frac{5}{16}\,k^2\,,
 \\
 c_2(k) &= \frac{1305}{512}-\frac{171}{64}\,k+\frac{343}{256}\,k^2+\frac{39}{64}\,k^3+\frac{25}{512}\,k^4\,,
 \\
 c_3(k) &= -\frac{8115}{2048}+\frac{255}{256}\,k-\frac{6795}{1024}\,k^2-\frac{585}{256}\,k^3-\frac{375}{2048}\,k^4\,.
\end{aligned}
\end{equation}
Using these results, it is straightforward to determine the post-freeze-out annihilation efficiency $J$ from eq.~\eqref{eq:J}.
We express it as
\begin{equation}
 J =
 \frac{A\,g_0(x_f)+B\,g_{-1}(x_f)+C\,g_{-2}(x_f)}{m_\chi^2}\,,
\end{equation}
where
\begin{equation}
 g_k(x_f)
 =
 \frac{2^{3+k}\,\Gamma(2+k/2)}{\sqrt{\pi}\,(3+k)}\,x_f^{-(3+k)/2}\left(1+\sum_{n=1}^\infty \frac{3+k}{3+k+2\,n}\frac{c_n(k)}{x_f^n}\right)\,.
\end{equation}

\subsection{$\Omega_0 h^2$ prediction}

\begin{figure}[t]
  \centering
  \includegraphics[width=.49\textwidth]{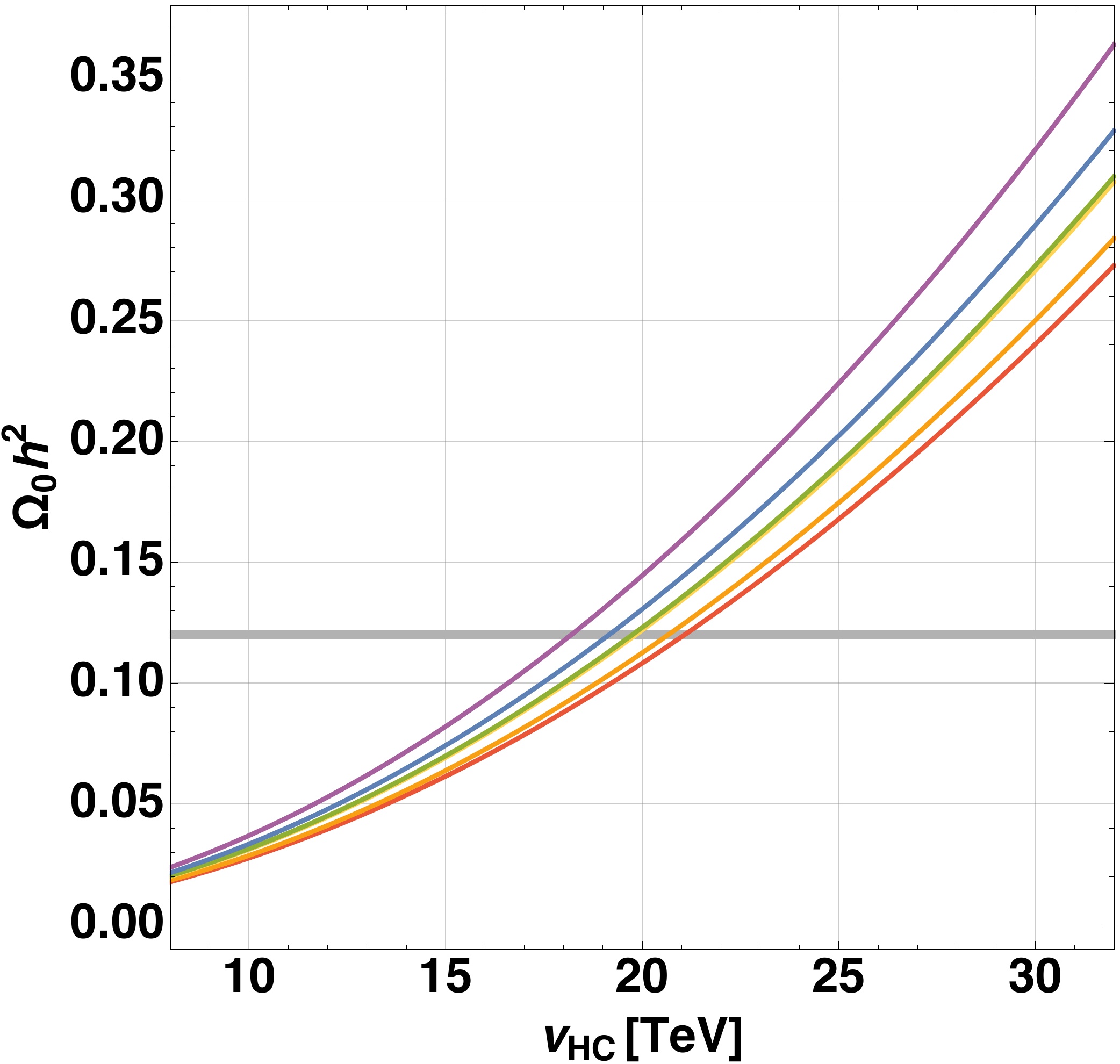}
  \hfill
  \includegraphics[width=.49\textwidth]{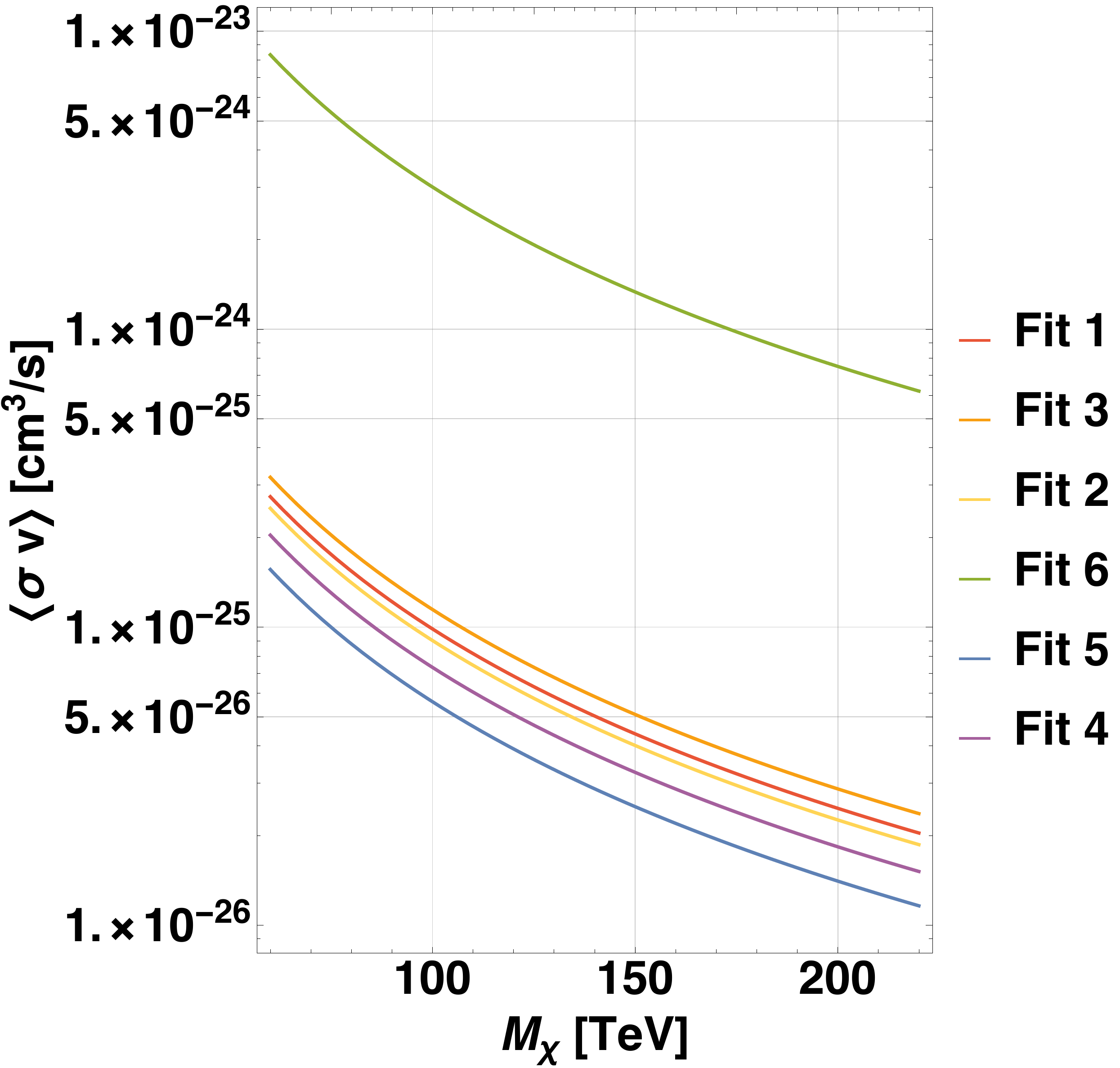}
  \caption{Left panel: Present DM relic density as a function of $v_{\HC}$ for the different fits in table~\ref{tab:sigma_ann} (see text for details). The thick horizontal gray line represents the measured value $\Omega_0 h^2 = 0.12$.
Right panel: estimate (see text for details) of the velocity-averaged DM-DM annihilation cross section into anything else, $\< \sigma_{\ann} v \>$, as a function of the DM mass, for the same fits as the left panel.}
  \label{fig:OmegaDM}
\end{figure}

We can now use the thermally averaged cross section in eq. (\ref{eq:sigmaeffv_ABCD}) to determine $\Omega_0 h^2$ via eqs. (\ref{eq:Tf}), (\ref{eq:J}) and (\ref{eq:Omega_DM}). It is interesting to note that ours is basically a one-parameter model, because the only quantity we can toggle is $v_{\HC}$ -- within the range in eq. (\ref{eq:vHC_range}), with $m_\chi \simeq 10 \, v_{\HC}$, see below eq. (\ref{eq:chipn}). The $\Omega_0 h^2$ prediction as a function of $v_{\HC}$ is presented in fig.~\ref{fig:OmegaDM} (left panel). The different lines correspond to an $\chi \chi$ annihilation cross section estimated from the different fits in table \ref{tab:sigma_ann}. Each line should be attached a theory error of around 10\% in the prediction of $\Omega_0 h^2$, due to the analytic procedure we used to estimate the relic density \cite{Griest:1990kh}. This error is not reported in the figure, to avoid clutter.
Quite remarkably, the measured value $\Omega_0 h^2 = 0.12$ is easily reproduced for $v_{\HC}$ in the range of eq. (\ref{eq:vHC_range}), for any of the fits in table \ref{tab:sigma_ann}. In particular, the $\Omega_0 h^2$ constraint tends to select $v_{\HC} \approx 20$ TeV and the spread of actual values due to the different fits is no larger than about 15\%.

\subsection{Comments on Dark-Matter indirect detection}

The annihilation cross section in eq. (\ref{eq:sigma_ann}) may also be used to place conservative upper bounds on the distortions in the measured cosmic-ray fluxes that our DM candidate may give rise to when annihilating.
Limits on the cross section $\sigma(\chi \chi \to X)$ for DM-pair annihilation to a given final state $X$ are usually given as $\< \sigma(\chi \chi \to X) v\>$, see e.g. \cite{Leane:2020liq}.
We have
\be
\label{eq:sigmavID}
\< \sigma(\chi \chi \to X) v\> ~\le~ \sum_X \< \sigma(\chi \chi \to X) v\> ~=~ \< \sigma_{\ann} v\> ~\approx~ \sigma_{\ann} \times \bar v~,
\ee
where in the last step we take $\bar v \sim 10^{-3}$ (see e.g. \cite{Cirelli:2010xx}).
In fig.~\ref{fig:OmegaDM} (right panel) we show the r.h.s. of eq.~(\ref{eq:sigmavID}) as a function of the DM mass, where we set $p_{\lab} = 10^{-3}\,m_\chi$.
We see that, with one exception, all the fits produce a velocity-averaged DM-DM annihilation ({\em total}) cross section below $3 \times 10^{-25}$ cm$^3$/s for $m_\chi > 70$ TeV (corresponding to $v_{\HC} > 7$ TeV).
This cross section is safely below the bounds in e.g.\ refs.~\cite{Abdallah:2018qtu,Lefranc:2016srp} for the same DM mass, while for higher DM masses the bounds are expected to be weaker and the prediction smaller.
The one exception is fit 6, whose prediction is a factor between about 30 and 50 larger than the other fits'.
However, none of the fits takes into account data for relative nucleon velocities as low as $10^{-3}$, which corresponds to $p_{\lab}=1$~MeV.
In fit~6, the lowest data points are given for a $[50,100]$~MeV bin.
More recent data is available from an experiment measuring $\bar{p}\,$-nucleus annihilation at $p_{\lab} = 15~{\rm MeV}$ \cite{Aghai-Khozani:2015pza}.
Under the assumption $\sigma_{\rm ann}^\text{$N\bar{p}$}\propto ZA^{1/3}$, valid at very low energies for the $\bar{p}\,$-nucleus interaction~\cite{Bianconi:2011zz,Batty:2000vr}, this data can be rescaled to the $p\bar{p}$ case.
Doing so, one finds that fit~6 clearly overshoots this data already at $p_{\lab} = 15~{\rm MeV}$, indicating that it may give much larger values for even smaller momenta.
Given also the conservative nature of our estimate via eq.~(\ref{eq:sigmavID}), i.e.\ the actual annihilation cross section in a given channel can be orders of magnitudes smaller than our estimated total annihilation cross section, we may conclude that our model is on the safe side of indirect-detection bounds.

This discussion underlines the importance of obtaining nucleon-antinucleon data at $p_{\lab}$ as small as 1~MeV, even though this might be very challenging. In fact, such data would provide very useful information for baryon-like DM scenarios.

\section{Conclusions}\label{sec:Conclusions}

We extend the SM by two flavors of massless vector-like fermions $\mc F$ charged under a new, `hypercolor' (HC), strong force.
An automatic outcome of this setup is the fact that the $\mc F$ fermions are endowed with an accidentally conserved hyperbaryon number, which makes the lightest hyperbaryons stable, and as such potential candidates for Dark Matter (DM).
We then gauge the $SU(2)_h$ diagonal subgroup of two global $SU(2)$ symmetries, one in the HC sector and the other one in the SM.
The former is the chiral symmetry $SU(2)_L^\mathcal{F}$ of left-handed hyperquarks, while the latter is the horizontal flavor symmetry of the two heavier generations of left-handed SM fermions.
This gauging has several desired consequences:
\begin{itemize}
 \item A mass term for the $SU(2)_h$ gauge bosons is generated by the chiral symmetry breaking in the HC sector and the corresponding NGBs become the longitudinal polarizations of the $SU(2)_h$ gauge bosons.
 \item The $SU(2)_h$ gauge bosons couple to the SM fermions in such a way that they can address experimental hints for lepton universality violation in $b \to s$ data, as advocated in~\cite{Guadagnoli:2018ojc}.
 \item The $SU(2)_h$ gauge bosons connect the baryon-like DM candidates in the HC sector to the SM fields, thereby playing the role of the DM mediator.
\end{itemize}
In order to generate the SM Higgs Yukawa couplings in the presence of the chiral $SU(2)_h$ gauge group, additional fermions $\Psi$ are introduced.
These fermions are vector-like under the SM gauge group but their right-handed components are charged under $SU(2)_h$.
They obtain masses from an `extended hypercolor' mechanism as the $\mc F$ fermions condense, they mix with the right-handed SM fermions, and they have a Higgs Yukawa coupling together with the left-handed SM fermions.
The final SM-like Higgs Yukawa couplings are then obtained from the mixing of right-handed SM fermions and $\Psi$ fermions.
All masses of beyond-the-SM fields only depend on one single scale, the $SU(2)_h$-symmetry breaking scale $v_{\HC}$, which is set by the $\mc F$-fermion condensate. It is expected to lie in the [10,30] TeV range by the mentioned $b \to s$ discrepancies.

We study the phenomenology of the hyperbaryon DM,
in particular its relic density, expected to be produced by a mechanism of thermal freeze-out, and its direct-detection signals. Because our model has, basically, one single scale, our DM phenomenology predictions are functions of this one parameter. We find remarkable that the $v_{\HC}$ range mentioned above yields the correct relic density for the lightest $\mc F$ hyperbaryons. In particular, the relic-density constraint selects $v_{\HC} \approx 20$~TeV, corresponding to a DM mass $m_\chi \simeq 10 \, v_{\HC} \approx $ 200~TeV. The stringent bounds from direct detection are quite comfortably satisfied. So our setup yields a composite DM candidate lying in the upper end of the mass range usually assumed for WIMP DM.

Taking advantage of nucleon-antinucleon annihilation data, we also provide a conservative estimate of indirect-detection signals of our scenario. In this context, we underline the importance of nucleon-antinucleon data at $p_{\lab}$ as small as 1~MeV, as they would offer important insights on baryon-like DM scenarios.

In conclusion, within our setup the DM in the Universe is generated by an inherently flavorful mechanism, a horizontal symmetry, and the DM mass scale is unambiguously related to the horizontal symmetry breaking scale. Flavor discrepancies provide a two-sided bound for this scale, and quite interestingly this very range produces a DM phenomenology in accord with observations. Although our setup is inspired by the current anomalies in $b \to s$ data,\footnote{For other studies of a possible connection between DM and current $B$-physics discrepancies see \cite{Sierra:2015fma,Belanger:2015nma,Allanach:2015gkd,Bauer:2015boy,Celis:2016ayl,Altmannshofer:2016jzy,Ko:2017quv,Ko:2017yrd,Cline:2017lvv,Sala:2017ihs,Ellis:2017nrp,Kawamura:2017ecz,Baek:2017sew,Cline:2017aed,Cline:2017qqu,Dhargyal:2017vcu,Chiang:2017zkh,Vicente:2018xbv,Falkowski:2018dsl,Arcadi:2018tly,Baek:2018aru,Azatov:2018kzb,Barman:2018jhz,Cerdeno:2019vpd,Trifinopoulos:2019lyo,DaRold:2019fiw,Fuentes-Martin:2019dxt,Han:2019diw,Guadagnoli:2020tlx,Huang:2020ris}.} the underlying idea may be easily adaptable to other collider discrepancies pointing to the tree-level exchange of a new gauge interaction with a symmetry-breaking scale in the tens-of-TeV range.

\section*{Acknowledgments}

DG warmly thanks Eugenio Del Nobile, Filippo Sala and Pasquale Dario Serpico for discussions on DM direct- and indirect-detection dynamics. Exchanges with Véronique Bernard are also acknowledged. This work is supported by an ANR PRC (contract n. 202650) and by the Labex Enigmass.

\appendix

\section{Mixing between SM and VL fermions}\label{sec:SM_VL_fermions}

For each $f\in\{u,d,e\}$, the $2\times 3$ matrices $\Delta_f$ in eq. (\ref{eq:LPsiQ}) can be singular-value decomposed as
\begin{equation}
 \Delta_f = U_L^{\Delta_f \dagger}\,\hat{\Delta}_f\,U_R^{\Delta_f}\,,
\end{equation}
where $U_L^{\Delta_f}$ and $U_R^{\Delta_f}$ are $2\times 2$ and $3\times 3$ unitary matrices, respectively, and $\hat{\Delta}_f$ is a $2\times 3$ rectangular diagonal matrix with two non-zero positive entries.
Without loss of generality, one can perform the redefinition
\begin{equation}
f^{\prime\,r}_R\to \left(U_R^{\Delta_f \dagger}\right)^{rk}\,f^{\prime\,k}_R\,,
\qquad
y_f^r \to y_f^k\,\left(U_R^{\Delta_f}\right)^{kr}\,,
\end{equation}
which entails that the mixing terms can be written as
\begin{equation}
 \mathcal{L}\supset
 - \Delta_f^{ij}\ \bar{\Psi}^{\prime\,f\,i}_L\, f^{\prime\,j}_R\,,
\end{equation}
where now $i,j\in\{2,3\}$, i.e.\ with a slight abuse of notation, $\Delta_f$ can be restricted to be a $2\times 2$ matrix given by
\begin{equation}
 \Delta_f = U_L^{\Delta_f \dagger}\,\hat{\Delta}_f\,,
\end{equation}
where $\hat{\Delta}_f$ is accordingly a $2\times 2$ diagonal matrix.

Once the hyperquarks $\mathcal{F}$ form the condensate
\begin{equation}
 \<\bar{\mathcal{F}}_L^\alpha \mathcal{F}_R^j\>
 =
 -\frac{1}{2}B_\mathcal{F}\,v_{\HC}^2\,\delta_{\alpha j}
 \approx
 -4\pi\,v_{\HC}^3\,\delta_{\alpha j}\,,
\end{equation}
where $\alpha$ is an index of the gauged $SU(2)_h$ and $j$ is an index of the global $SU(2)_R^\mathcal{F}$, the four-fermion operators in eq.~\eqref{eq:EHC} yield the mass terms
\begin{equation}
 -m_\Psi^{\prime\,i \alpha}\,(\bar{\Psi}_L^i \Psi_R^\alpha)+h.c.\,,
 \qquad
 \text{where}
 \qquad
 m_\Psi^{\prime\,i \alpha} = \frac{c_{i \alpha}+\tilde{c}_{i \alpha}}{2}\,B_{\mathcal{F}}\,\frac{v_{\HC}^2}{\Lambda_{\EHC}^2}\,.
\end{equation}
Consequently, for each $f\in\{u,d,e\}$ there is a $2\times 4$ mass-mixing matrix
\begin{equation}
 M_f^\prime =
 \begin{pmatrix}
  \Delta_f
  &
  m_{\Psi^f}^\prime
 \end{pmatrix},
\end{equation}
such that
\begin{equation}
 \mathcal{L}\supset
 -\bar{\Psi}_L^{\prime\,f}\,M^\prime_f\,
 \begin{pmatrix}
  f_R^\prime \\
  \Psi_R^{\prime\,f}
 \end{pmatrix}\,.
\end{equation}
The singular value decomposition of the matrix $M_f^\prime$ yields
\begin{equation}
 M_f =
 \begin{pmatrix}
  0
  &
  m_{\Psi^f}
 \end{pmatrix}
  =
 U^{f\,\dagger}_L\,
 M_f^\prime\,
 \begin{pmatrix}
  U_R^{f^\prime f} &
  U_R^{f^\prime \Psi^f}
  \\
  U_R^{\Psi^{\prime f} f} &
  U_R^{\Psi^{\prime f} \Psi^f}
 \end{pmatrix}\,,
\end{equation}
where $U_L$ is a $2\times 2$ unitary matrix, the $U_R$ are $2\times 2$ submatrices of a $4\times 4$ unitary matrix, and $m_{\Psi^f}$ is a $2\times 2$ diagonal matrix with real positive entries.
Using these unitary matrices to transform the fields to the mass eigenbasis (before EW symmetry breaking), the SM Yukawa matrices read
\begin{equation}
 Y_f
 =
 \begin{pmatrix}
  y_f^1
  &
  \begin{pmatrix}
  y_{f}^2 & y_{f}^3
 \end{pmatrix}
 \,U_R^{f^\prime f}
  \\
  0_{2\times 1} & y_F\,U_R^{\Psi^{\prime f} f}
 \end{pmatrix}\,.
\end{equation}

\bibliographystyle{JHEP}
\bibliography{bibliography}

\end{document}